\documentclass[onecolumn,superscriptaddress,
		aps,prx,floatfix]{revtex4-2}

\usepackage[normalem]{ulem}
\usepackage{graphicx}
\usepackage{bm}
\usepackage{color}
\usepackage{epstopdf}
\usepackage{amsmath}
\usepackage{amssymb}
\usepackage{epstopdf}
\usepackage{lipsum}

\usepackage{xfrac}

\usepackage[T1]{fontenc} % polskie ogonki

\usepackage{float}

\usepackage{tikz}
\newcommand*\circled[1]{\tikz[baseline=(char.base)]{
            \node[shape=circle,draw,inner sep=2pt] (char) {#1};}}

\usepackage{multirow}

\usepackage[urlcolor=blue,colorlinks=true,citecolor=blue,
linkcolor=blue,pdfstartview={FitH},bookmarks=false]{hyperref}

\usepackage{xcolor}
\definecolor{mygreen}{rgb}{0.0, 0.6, 0.0}
\definecolor{pjorange}{rgb}{0.8, 0.3, 0.0}
\definecolor{jlblue}{rgb}{0.2, 0.5, 0.7}
\newcommand{\AP}[1]{\textcolor{black}{#1}}

\graphicspath{{fig/}{./fig/}{.}}

\begin{document}

\title{Chiral phonons in binary compounds $A$Bi ($A$ = K, Rb, Cs) with P2$_1$/c structure}

\author{Jakub Sk\'{o}rka}
\email[e-mail: ]{js2368@cantab.ac.uk}
\affiliation{Institute of Nuclear Physics, Polish Academy of Sciences, 
ul. W. E. Radzikowskiego 152, 31342 Krak\'{o}w, Poland}

\author{Konrad J. Kapcia}
\email[e-mail: ]{konrad.kapcia@amu.edu.pl}
\affiliation{\mbox{Institute of Spintronics and Quantum Information, Faculty of Physics,} Adam Mickiewicz University in Pozna\'n, 
ul. Uniwersytetu Pozna\'{n}skiego 2, 61614 Pozna\'{n}, Poland}
\affiliation{\mbox{Center for Free-Electron Laser Science CFEL, Deutsches Elektronen-Synchrotron DESY}, Notkestr. 85, 22607 Hamburg, Germany}

\author{Pawe\l{} T. Jochym}
\email[e-mail: ]{pawel.jochym@ifj.edu.pl}
\affiliation{Institute of Nuclear Physics, Polish Academy of Sciences, 
ul. W. E. Radzikowskiego 152, 31342 Krak\'{o}w, Poland}

\author{Andrzej Ptok}
\email[e-mail: ]{aptok@mmj.pl}
\affiliation{Institute of Nuclear Physics, Polish Academy of Sciences, 
ul. W. E. Radzikowskiego 152, 31342 Krak\'{o}w, Poland}

\date{\today}

\begin{abstract}
Binary compounds $A$Bi ($A$ = K, Rb, Cs) crystallize in P2$_1$/c structure containing both clockwise and anticlockwise chiral chains of Bi atoms.
Electronic band structure exhibits the insulating nature of these systems, with the band gap about $0.25$~eV.
The presented study of dynamical properties confirm a stability of the system with P2$_1$/c symmetry.
The crystal structure contains the quasi-one-dimensional Bi chains, exhibiting four-fold-like rotational ``local'' symmetry.
Nevertheless, the system formally posses two-fold rotational symmetry.
Independently of the absence of the three-fold (or higher) rotational symmetry axes for the whole crystal, the chiral modes propagate along the Bi atom chains in these systems.
We discuss basic properties of these modes in monoatomic chiral chains.
We show that the two-fold rotational symmetry axis affects the main properties of the chiral phonons, which are not realized at the high-symmetry points, but along some paths between them in the reciprocal space.
\AP{In addition, in the doped system, the chiral phonons possess non-zero total angular momentum.}
\end{abstract}

\maketitle


\section{Introduction}

Chirality plays a fundamental role affecting system properties.
One of the well-known examples of such a role are chiral molecules absorbing different amounts of left-handed (LHP) or right-handed (RHP) polarized light~\cite{kneer.roller.18}.
Such a distinguishing difference of interplay of the system with LHP or RHP light allows for spectroscopic characterization of the system's chirality.
However, the concept of the chirality can be also applied to the collective excitation in a system, such as phonons~\cite{chen.zhang.18}.

\AP{Similarly like the electronic systems, the phononic systems can possess the topological features~\cite{liu.chen.20,li.liu.21,wang.yang.22}, which can be exhibited in the form of nodal lines~\cite{li.wang.20,zheng.xia.20,chen.xie.22}, Dirac and Weyl points~\cite{liu.qian.20,zhong.liu.21,wang.zhou.22,wang.zhou.22b}, surface states~\cite{zhang.miao.19,liu.huang.22,basak.kobialka.23}, and many others.}
The chiral phonons can be realized in a large class of the two-dimensional (2D) systems with hexagonal symmetry~\cite{chen.zhang.18}, like the honeycomb lattice~\cite{zhang.niu.15,liu.lian.17} or the kagome lattice~\cite{chen.wu.19}.
Indeed, recently the observation of the chiral phonons in layered structures was reported in, e.g., B-decorated graphene~\cite{gao.zhang.18}, dichalcogenide monolayers~\cite{chen.zheng.15,zhu.yi.18,li.wang.19,li.wang.19b,liu.vanbaren.19} and bilayers~\cite{delhomme.vaclavkova.20,zhang.srivastava.20}, moir\'{e} twisted bilayer~\cite{suri.wang.21,maity.mostofi.22}, or CrBr$_{3}$~\cite{yin.ulman.21}.
However, the three-dimensional (3D) systems can also exhibit the chiral phonons, e.g., cuprates~\cite{grissonnanche.theiault.20}, perovskites~\cite{nova.cartella.17,juraschek.fechner.17,jurascheek.spaldin.19,li.fauque.20} and their superlattices~\cite{jeong.kim.22}, \AP{compounds relized the Laves phase~\cite{basak.piekarz.22}}, Fe$_{3}$GeTe$_{2}$~\cite{du.tang.19}, \AP{{\it shandite}-like compounds~\cite{basak.kobialka.23}}, magnetic topological insularotrs \AP{$T$Bi$_{2}$Se$_{4}$ ($T$=Mn,Fe)}~\cite{kobialka.sternik.22}, or CoSn-like compounds~\cite{ptok.kobialka.21}.

The chiral phonons can be induced by light in some materials~\cite{nova.cartella.17,juraschek.fechner.17,chen.lu.19,jurascheek.spaldin.19,juraschek.narang.20}, which shows their interplay with circularly polarized light.
Additionally,  the effects of the magnetic field on the chiral phonons are well-known~\cite{cheng.schumann.20}.
Firstly, the chiral modes can be induced by the external magnetic field, as a consequence of decoupling of the degenerated mode at the $\Gamma$ point~\cite{schaack.76,thalmeier.fulde.77,li.wang.19,li.wang.19b,liu.vanbaren.19,juraschek.neuman.22}.
Secondly, the presence of the magnetic field can lead to phonon Hall effect~\cite{strohm.rikken.05,sheng.sheng.06,inyushkin.taldenkov.07,zhang.ren.11,zhang.niu.14,zhang.niu.15} and some contribution to the thermal Hall effect due to the phonon-magnon coupling~\cite{sugii.shimozawa.17,kasahara.sugii.18,grissonnanche.legros.19,zhang.zhang.19,li.fauque.20,boulanger.grissonnanche.20,yokoi.ma.21,zhang.xu.21,boulanger.grissonnanche.21}.

\begin{figure}[!b]
\centering
\includegraphics[width=0.5\linewidth]{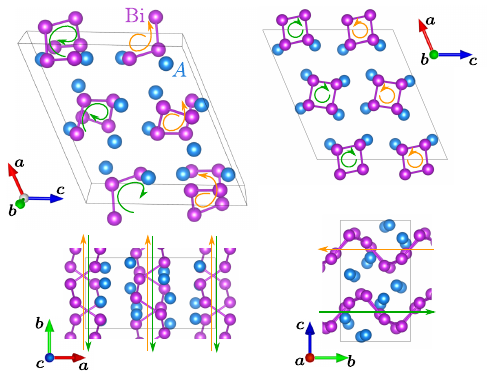}
\caption{
Crystal structure of $A$Bi ($A=$K, Rb, Cs) with P2$_{1}$/c symmetry.
The green/orange arrows indicate clockwise/anticlockwise chirality of the Bi chains.
\label{fig.schemat}
}
\end{figure}

\paragraph*{Motivation ---}
Typically, the realization of the chiral phonon modes is discussed in the context of the system exhibiting the hexagonal (three-fold or six-fold rotational) symmetry, which is one of the conditions for a realization of these modes~\cite{coh.19}.
Recent experimental study of Ishito {\it et~al.}~\cite{ishito.mao.21} shows existence of the chiral phonons in the 3D chiral binary crystal of $\alpha$-HgS.
In this case, the chiral phonons are possessed due to the mass imbalance of the atoms in the chiral chains.
This translation symmetry breaking as a source of the chiral phonons was discussed previously in the context of the 2D honeycomb lattice~\cite{zhang.niu.15,xu.chen.18}.
However, the chiral crystals have natural tendency to a realization of the chiral phonons. 
In the recent study of Wang {\it et~al.}~\cite{wang.li.21}  the chiral phonons are investigated in systems with four-fold rotational symmetry, i.e., in 2D MnAs monolayer and 3D tetragonal $\alpha$-cristobalite (SiO$_{4}$) chiral crystal.
In these cases, the chiral phonons were realized in the system with four-fold rotation symmetry. 
Additionally, $\alpha$-SiO$_{4}$ belongs to the nonsymmorphic group (P4$_{1}$2$_{1}$2 or P4$_{3}$2$_{1}$2).
This system is also an example of the nonsymmorphic system realizing the chiral modes~\cite{zhang.murakami.21}.

Here, we focus on the recently discovered binary compounds $A$Bi ($A=$K, Rb, Cs)~\cite{emmerling.langin.04}.
For light alkali metals ($A=$Li, Na), this type of compounds crystallize in tetragonal P4/mmm phase, and they can exhibit superconducting properties~\cite{gornicka.gutowksa.20,kushwaha.krizan.14,li.cheng.15}.
In opposition to this, the systems containing heavier alkali metals ($A=$K, Rb, Cs) crystallize in monoclinic P2$_{1}$/c structure (cf. Fig.~\ref{fig.schemat}).
This structure, similarly to $\alpha$-HgS, contains the chiral chains formed by Bi atoms (which posses four-fold-like ``local'' rotational symmetry).
However, unlike the $\alpha$-HgS, the system does not exhibit three-fold helical symmetry.
In fact,  in the case of $A$Bi$_{2}$ with P2$_{1}$/c symmetry, only the two-fold rotation symmetry is formally realized, which creates great opportunity to study the chiral phonons in 3D system possessing nonsymmporphic group~\cite{zhang.murakami.21} and containing the monoatomic chiral chains~\cite{chen.wu.21} in the absence of hexagonal symmetry~\cite{wang.li.21}.

The paper is organized as follows.
First, we shortly describe details of the used theoretical methods (Sec.~\ref{sec.calc}).
Next, in Sec.~\ref{sec.basic}, we present the basic properties of the studied systems.
Particularly, we investigate the crystal structure parameters (Sec.~\ref{sec.struc}) and the electronic band structure (Sec.~\ref{sec.el_band}).
Finally, we analyze the dynamical properties of the system (Sec.~\ref{sec.dynamical}), in the context of system stability (Sec.~\ref{sec.stability}), Raman spectroscopy (Sec.~\ref{sec.raman}), and appearance of the chiral phonons.
Here, we are focused on the chiral phonons showing up in the clean system (Sec.~\ref{sec.chiral_clean}).
Next, we discuss the possibility of realization of the chiral modes in the presence of impurities (Sec.~\ref{sec.chiral_imp}), which leads to chiral modes with total non-zero pseudoangular momentum (PAM).
Finally, we discuss the propagation of the chiral modes along the Bi atoms chains (Sec.~\ref{sec.chiral_prop}).
We conclude our results in Sec.~\ref{sec.summary}.

\section{Calculation Details}
\label{sec.calc}

\subsection{Numerical calculations}

The first-principles calculations were performed using the projector augmented wave (PAW) potentials~\cite{blochl.94} and the generalized gradient approximation (GGA) in the Pardew, Burke, and Ernzerhof (PBE) parametrization~\cite{perdew.burke.96} implemented in the {\sc VASP} code~\cite{kresse.furthmuller.96,kresse.hafner.94,kresse.houbert.99}.
The calculations were performed with and without the spin--orbit coupling (SOC).
For the summation over the reciprocal space, we used an $8 \times 16 \times 8$  Monkhorst-Pack ${\bm k}$-point grid~\cite{monkhorst.pack.76}.
The energy cutoff for plane-wave expansion was set to $350$~eV.
The crystal structures were optimized using the conjugate gradient (CG) 
technique with the energy convergence criteria set at $10^{-8}$ and $10^{-6}$ for electronic and ionic iterations, respectively.
The systems symmetries were evaluated within the {\sc FindSym}~\cite{stokes.hatch.05} and {\sc Spglib}~\cite{togo.tanaka.18} packages, while the reciprocal space analyses were done within {\sc SeekPath}~\cite{hinuma.pizzi.17}.

The dynamical properties of the systems were calculated within the Parlinski--Li--Kawazoe {\it direct method}~\cite{parlinski.li.97} implemented in the {\sc Phonopy} software~\cite{phonopy}.
In this approach, atoms are displaced from their equilibrium positions and  the Hellmann--Feynman forces acting on all atoms in the non-equivalent positions in the supercell are calculated.
Next, the interatomic force constants (IFC) are determined.
Additionally, the systems stability was examined within {\sc alamode} package~\cite{tadano.gohda.14,tadano.tsuneyuki.18}.
Here, the phonon calculations were performed within the conventional cell.

\subsection{Dynamical matrix}

Lattice dynamics of the system is described by the dynamical matrix:
\begin{eqnarray}
\nonumber D_{\alpha\beta}^{jj'} \left( {\bm q} \right) = \frac{1}{\sqrt{m_{j}m_{j'}}} \sum_{n} \Phi_{\alpha\beta} \left( 0j, nj' \right) \exp \left( i {\bm q} \cdot {\bm R}_{j'n} \right) , \\
\label{eq.dynmat}
\end{eqnarray}
where ${\bm q}$ is the phonon wave vector and $m_{j}$ denotes the mass of $j$-th atom. 
Here, $\Phi_{\alpha\beta} \left( 0j, nj' \right)$ is the IFC tensor between $j$-th and $j'$-th atoms located in the initial ($0$) and $n$-th primitive unit cell ($\alpha$ and $\beta$ denotes direction index $x$, $y$, or $z$).
The phonon spectrum for  wave vector ${\bm q}$ is given by eigenproblem of the dynamical matrix (\ref{eq.dynmat}):
\begin{eqnarray}
\omega_{\varepsilon{\bm q}}^{2} \text{e}_{\varepsilon{\bm q}\alpha j} = \sum_{j', \beta} D_{\alpha\beta}^{jj'} \left( {\bm q} \right) \text{e}_{\varepsilon{\bm q}\beta j'} .
\end{eqnarray}
For each wave vector ${\bm q}$ there are $3N$ branches (where $N$ is a number of the atoms in the primitive unit cell), with frequencies $\omega_{\varepsilon{\bm q}}$.
Moreover, the selected mode is described by polarization (eigen)vectors $\text{e}_{\varepsilon{\bm q}\alpha j}$, which components denote displacement of the $j$-th atom along $\alpha$-th direction.

\subsection{Chirality of phonons}
\label{sec.chiral_teo}

The phonon chirality can be characterized by the polarization of phonons, which comes from the circular vibration of sublattices~\cite{zhang.niu.15}.
We are particularly interested in the study of the chiral phonons realized by the Bi atoms chains and propagating along the ${\bm b} \parallel y$ direction [see Fig.~\ref{fig.schemat}(a)].
To investigate these phonons, let us introduce the left-handed and right-handed circular polarizations in the $xz$-plane (perpendicular to ${\bm b}$).
In this case, we construct a new basis defined as 
\mbox{$\vert Y_{1} \rangle = \left( 0 \; 1 \; 0 \; \cdots \right)$};
\mbox{$\vert R_{1} \rangle = \sfrac{1}{\sqrt{2}} \left( 1 \; 0 \; i \; \cdots \right)$};
\mbox{$\vert L_{1} \rangle = \sfrac{1}{\sqrt{2}} \left( 1 \; 0 \; -i \; \cdots \right)$};
$\cdots$;
\mbox{$\vert Y_{j} \rangle = \left( \cdots \; 0 \; 1 \; 0 \; \cdots \right)$};
\mbox{$\vert R_{j} \rangle = \sfrac{1}{\sqrt{2}} \left( \cdots \; 1 \; 0 \; i \; \cdots \right)$}; and 
\mbox{$\vert L_{j} \rangle = \sfrac{1}{\sqrt{2}} \left( \cdots \; 1 \; 0 \; -i \; \cdots \right)$};
i.e., two in-$xz$-plane components of polarization vector for $j$-th atom are replaced by the Jones vectors coefficients, while the out-of-$xz$-plane component is unchanged.
In the new basis each polarization vector $\text{e} = \text{e}_{\varepsilon{\bm q}\alpha j}$ is represented in the form:
\begin{eqnarray}
\text{e} = \sum_{j} \left( \alpha_{j}^{Y} \vert Y_{j} \rangle + \alpha_{j}^{R} \vert R_{j} \rangle + \alpha_{L}^{j} \vert L_{j} \rangle \right) ,
\end{eqnarray}
where $\alpha_{j}^{V} = \langle V_{j} \vert \text{e} \rangle$, for $V \in \{ Y, R, L \}$ and $j \in \{ 1,2,\cdots,N \}$.
Then, the operator for phonon circular polarization along $y$ axis can be defined as:
\begin{eqnarray}
\hat{S}^{y} \equiv \sum_{j=1}^{N} s_{j}^{y} = \sum_{j=1}^{N} \left( | R_{j} \rangle \langle R_{j} | + | L_{j} \rangle \langle L_{j} | \right) , 
\end{eqnarray}
and the phonon circular polarization is equal to
\begin{eqnarray}
s^{y}_\text{ph} = \text{e}^{\dagger} \hat{S}^{y} \text{e} = \sum_{j=1}^{N} s_{j}^{y} \hslash = \sum_{j=1}^{N} \left( | \alpha_{j}^{R} |^{2} - | \alpha_{j}^{L} |^{2} \right) \hslash .
\end{eqnarray}
Since $\sum_{j} \left( | \alpha_{j}^{R} |^{2} + | \alpha_{j}^{L} |^{2} \right) = 1$, the phonon circular polarization fulfill the condition $| s^{y}_\text{ph} | \leq \hslash$.
The $s^{y}_\text{ph}$ has the same form with that of phonon angular momentum along $y$ axis~\cite{zhang.niu.14}, while the phonon ciruclar polarization is equivalent to the phonon angular momentum.

Here, we introduce $s_{j}^{y}$, which denotes the contribution of each atom to the phonon circular polarization.
When for some atom $s_{j}^{y}$ is non-zero, then the circular motion around the equilibrium positions occurs and chiral phonons are present in the system.
Thus, the total phonon circular polarization can be related to the total PAM of the system, which will be more elaborated in Sec.~\ref{sec.chiral_clean}.
For a realization of the non-zero PAM several conditions should be fulfilled, e.g., breaking of the inversion symmetry and/or time reversal symmetry~\cite{coh.19}.
Nevertheless, also in the system with the conserved inversion symmetry and the time reversal symmetry, the chiral phonons can be realized, while the PAM is equal zero.
Example of this case is the ideal honeycomb lattice (containing two non-equivalent atomic positions: A and B) with two identical atoms in both non-equivalent positions.
Such situation occurs in the honeycomb sublattice of the $p$-block atoms in the CoSn-like compounds~\cite{ptok.kobialka.21}.
In these systems, the non-zero PAM can be ``induced'' by the breaking of the spatial inversion symmetry, e.g., by different masses of atoms in A and B positions~\cite{zhang.niu.15}.

\begin{table}[!t]
\caption{
\label{tab.dft}
Comparison between theoretical and experimental crystal parameters of $A$Bi ($A=$K, Rb, and Cs).
Theoretical calculations are performed in the presence of the spin--orbit coupling.}
%\begin{ruledtabular}
\begin{tabular}{ccc}
 & Theory (this paper) & Experiment (Ref.~\cite{emmerling.langin.04}) \\
\hline
\multicolumn{3}{c}{\bf KBi} \\
\hline
$a$ & $14.640$~\AA\ & $14.223$~\AA\ \\
$b$ & $7.437$~\AA\ & $7.248$~\AA\ \\
$c$ & $13.756$~\AA\ & $13.420$~\AA\ \\
$\beta$ & $113.4$ & $113.0$ \\
\hline 
K(1) & ($0.1210$, $0.1156$, $0.1474$) & ($0.1194$, $0.1180$, $0.1472$) \\ 
K(2) & ($0.3956$, $0.3035$, $0.0370$) & ($0.3946$, $0.3005$, $0.0391$) \\
K(3) & ($0.6547$, $0.0745$, $0.1934$) & ($0.6537$, $0.0714$, $0.1911$) \\
K(4) & ($0.8497$, $0.3571$, $0.0532$) & ($0.8482$, $0.3608$, $0.0538$) \\
Bi(1) & ($0.0731$, $0.6272$, $0.1831$) & ($0.0760$, $0.6290$, $0.1819$) \\ 
Bi(2) & ($0.1101$, $0.3700$, $0.3702$) & ($0.1143$, $0.3715$, $0.3751$) \\
Bi(3) & ($0.3931$, $0.0652$, $0.2709$) & ($0.3908$, $0.0652$, $0.2746$) \\
Bi(4) & ($0.5770$, $0.2979$, $0.3874$) & ($0.5818$, $0.2976$, $0.3910$) \\
\hline
\hline
\multicolumn{3}{c}{\bf RbBi} \\
\hline
$a$ & $15.2427$~\AA\ & $14.742$~\AA\ \\
$b$ & $7.656$~\AA\ & $7.502$~\AA\ \\
$c$ & $14.288$~\AA\ & $13.921$~\AA \\
$\beta$ & $113.4$ &  $113.0$ \\
\hline 
Rb(1) & ($0.1282$, $0.1105$, $0.1521$) & ($0.1517$, $0.1202$, $0.1443$) \\ 
Rb(2) & ($0.3915$, $0.3102$, $0.0337$) & ($0.3940$, $0.3015$, $0.0348$) \\
Rb(3) & ($0.6533$, $0.0811$, $0.1868$) & ($0.6551$, $0.0737$, $0.1881$) \\
Rb(4) & ($0.8569$, $0.3519$, $0.0463$) & ($0.8489$, $0.3600$, $0.0511$) \\
Bi(1) & ($0.0752$, $0.6231$, $0.1937$) & ($0.0726$, $0.6337$, $0.1845$) \\ 
Bi(2) & ($0.1003$, $0.3621$, $0.3665$) & ($0.1086$, $0.3729$, $0.3667$) \\
Bi(3) & ($0.4009$, $0.0708$, $0.2714$) & ($0.3959$, $0.0662$, $0.2721$) \\
Bi(4) & ($0.5753$, $0.3039$, $0.3800$) & ($0.5779$, $0.2988$, $0.3844$) \\
\hline
\hline
\multicolumn{3}{c}{\bf CsBi} \\
\hline
$a$ & $15.841$~\AA\ & $15.237$~\AA\ \\
$b$ & $7.845$~\AA\ & $7.737$~\AA\ \\
$c$ & $14.809$~\AA\ & $14.399$~\AA\ \\
$\beta$ & $112.8$ & $112.7$ \\
\hline 
Cs(1) & ($0.1329$, $0.1072$, $0.15401$) & ($0.1505$, $0.1292$, $0.1344$) \\ 
Cs(2) & ($0.3905$, $0.3116$, $0.03311$) & ($0.3947$, $0.2997$, $0.0322$) \\
Cs(3) & ($0.6530$, $0.0814$, $0.18288$) & ($0.6588$, $0.0731$, $0.1877$) \\
Cs(4) & ($0.8620$, $0.3492$, $0.04381$) & ($0.8452$, $0.3671$, $0.1877$) \\
Bi(1) & ($0.0751$, $0.6188$, $0.20059$) & ($0.0664$, $0.6472$, $0.1817$) \\ 
Bi(2) & ($0.0914$, $0.3576$, $0.36181$) & ($0.10429$, $0.3855$, $0.3553$) \\
Bi(3) & ($0.4074$, $0.0711$, $0.27253$) & ($0.3995$, $0.06275$, $0.2697$) \\
Bi(4) & ($0.5724$, $0.3064$, $0.37272$) & ($0.5732$, $0.2974$, $0.3778$) 
\end{tabular}
%\end{ruledtabular}
\end{table}

\begin{figure*}
\centering
\includegraphics[width=\linewidth]{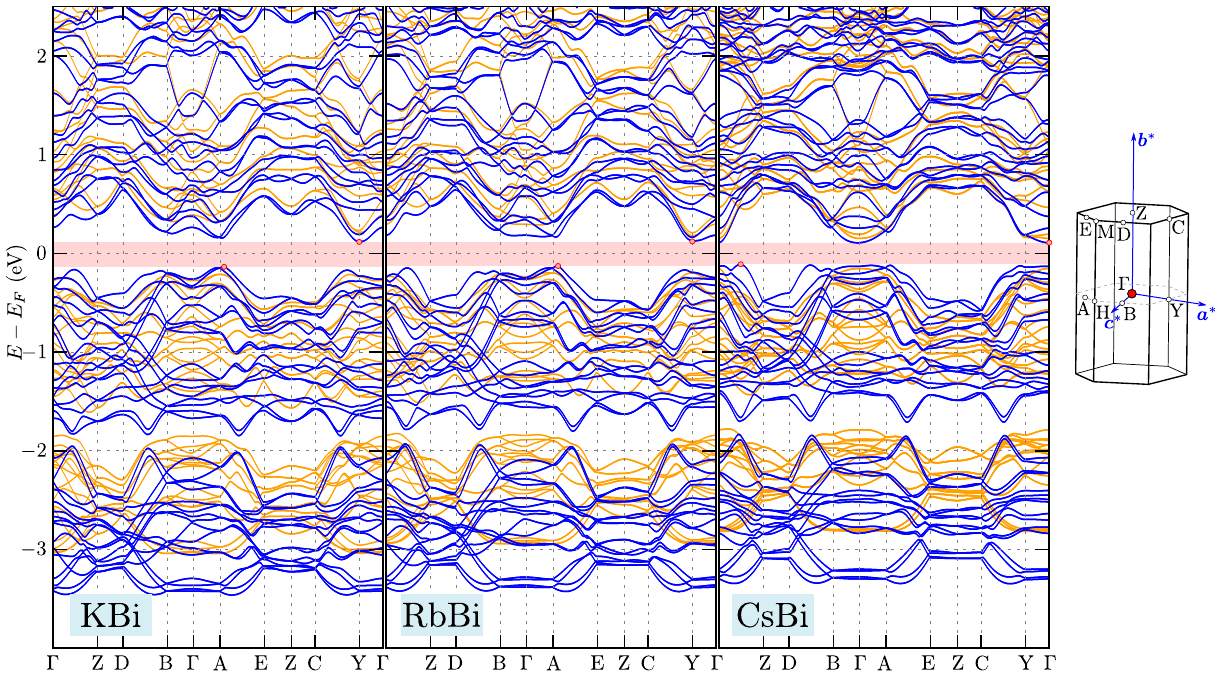}
\caption{
Electronic band structure of $A$Bi ($A=$K, Rb, Cs; from left to right, respectively) in the absence and presence of the spin--orbit coupling (orange and blue lines, respectively).
Red points mark positions of the bottom (top) of the conduction (valence) band (in the case with spin--orbit coupling).
On the right side of the figure, the first Brillouin zone and the high symmetry points of P2$_{1}$/c space group are presented.
\label{fig.el_band}
}
\end{figure*}

\section{Basic properties}
\label{sec.basic}

\subsection{Crystal structures}
\label{sec.struc}

The $A$Bi ($A=$K, Rb, and Cs) crystallize in the P2$_{1}$/c symmetry (nonsymmorphic space group No.~14 with point group $C_{2h}^{-5}$).
These systems exhibit the inversion symmetry $\{ -1 | 0 \}$, mirror symmetry $\{ m_{010} | 0 \; \sfrac{1}{2} \; \sfrac{1}{2} \}$, and a two-fold rotational symmetry $\{ 2_{010} | 0 \; \sfrac{1}{2} \; \sfrac{1}{2} \}$ (i.e., around the $y$ direction, which is parallel to the lattice vector ${\bm b}$ in Fig.~\ref{fig.schemat}).
Positions of both atoms are given by $4e$ Wyckoff positions.
A comparison between calculated and experimental crystal lattice parameters is presented in Tab.~\ref{tab.dft}.
The found lattice constants as well as the atomic positions are close to these reported experimentally~\cite{emmerling.langin.04}.

\subsection{Electronic band structure}
\label{sec.el_band}

The band structures of $A$Bi ($A=$K, Rb, Cs) presented in Fig.~\ref{fig.el_band} clearly show that the investigated materials are insulators~\cite{emmerling.langin.04}.
The insulating gap (marked by the red band) is approximately equal to $0.25$~eV for all compounds, and it has indirect character (the bottom of the conduction band and top of the valence bands are marked by red points).

As we can see, the band structure is constructed by the folded-like bands due to the existence of the chiral chains of  Bi atoms.
Additionally, the spin--orbit coupling leads to strong modification of the band structure (cf. orange and blue lines in Fig.~\ref{fig.el_band}).
This is expected in the presence of Bi atoms, which have large impact on effects of the spin--orbit coupling in similar systems~\cite{shao.luo.16,golab.wiendlocha.19}.
Indeed, the topological insulating state was reported in LaBi~\cite{lou.fu.17} or CeBi~\cite{huan.shi.21} (both with Fm$\bar{3}$m symmetry).
Here, the spin--orbit coupling modifies the band structure in two ways.
Firstly, it leads to a typical lift of degeneracy and splitting of the bands.
The second effect is more unexpected, as it leads to an increase of the bandwidths of the separated bands below the Fermi level (it is well visible for bands in a range of energies from $-3.5$~eV to $-2.0$~eV).
Moreover, the spin--orbit coupling strongly affects positions of the bottom (top) of the conduction (valence) band.

\paragraph*{The Dirac nodal lines ---}
The time reversal symmetry is still preserved in investigated systems due to the absence of the magnetic order. 
Additionally, the inversion symmetry $\mathcal{I}$ is guaranteed by the space group P2$_{1}$/c.
These conditions allow for occurrence of the nodal lines in the presence of the weak spin--orbit coupling~\cite{kim.wieder.15,yu.weng.15,chan.chiu.16}.
Due to the symmetry of the system (for detailed discussion, see Ref.~\cite{geilhufe.bouhon.17}), the topologically protected Dirac nodal lines are found in the ${\bm a}^{\ast}{\bm c}^{\ast}$ plane.
However, the presence of the heavy Bi atoms~\cite{shanavas.popovic.14} leads to relatively strong spin--orbit coupling.
This could be the reason for the absence of the Dirac nodal line in the band structure.
Nevertheless, the highly degenerated points in the band stricture are still present in some high symmetry points as a consequence of the crystal symmetry.

%%%%%%%%%%%%%%%%%%%%%%%%%%%%%%%%%%%%%%%
%%%%%%%%%%%%%%%%%%%%%%%%%%%%%%%%%%%%%%%
%%%%%%%%%%%%%%%%%%%%%%%%%%%%%%%%%%%%%%%

\section{Dynamical properties}
\label{sec.dynamical}

\subsection{System stability and phonon dispersion}
\label{sec.stability}

\begin{figure}[!pt]
\centering
\includegraphics[width=0.57\linewidth]{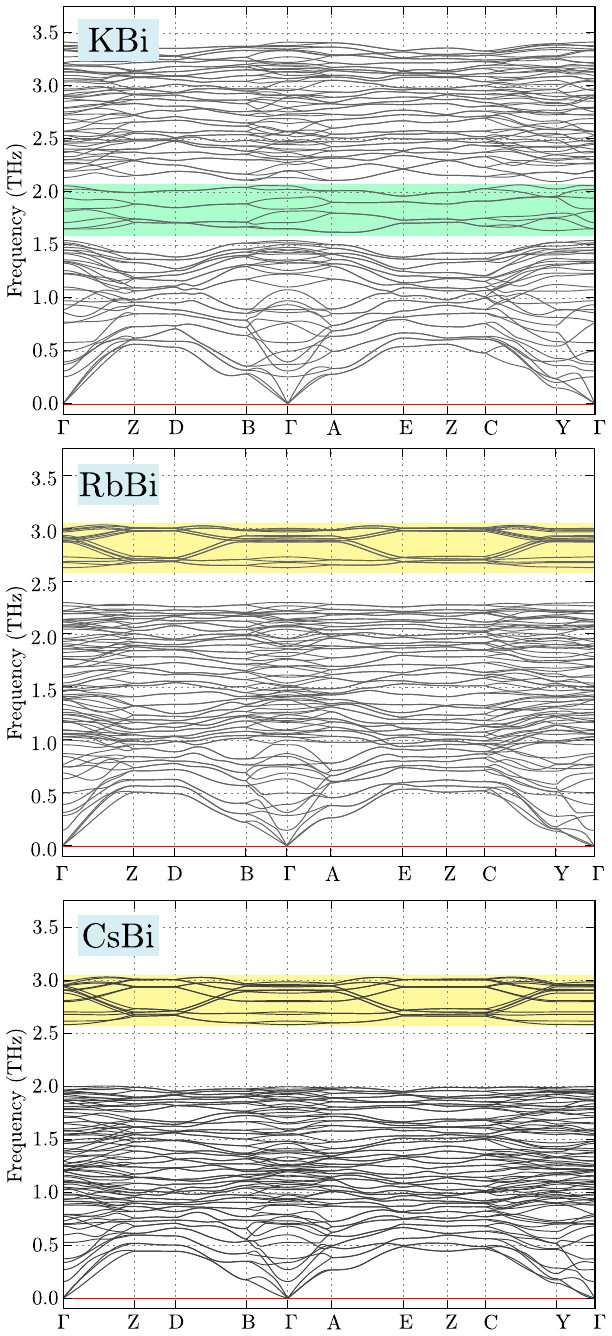}
\caption{
Phonon dispersion curves of $A$Bi ($A=$K, Rb, Cs; from the top) with P2$_{1}$/c symmetry.
\label{fig.ph_band}
}
\end{figure}

\begin{figure}[!t]
\centering
\includegraphics[width=0.5\linewidth]{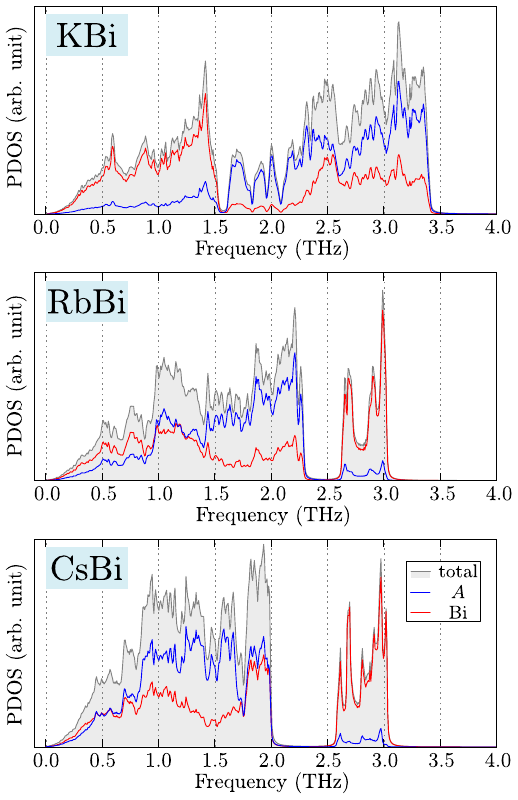}
\caption{
The phonon density of states of $A$Bi ($A=$K, Rb, Cs; from the top) with P2$_{1}$/c symmetry.
The total density of states is presented by grey color (shadow), while the partial density of states of $A$ and Bi are denoted by blue and red lines, respectively.
\label{fig.ph_dos}
}
\end{figure}

The calculated phonon dispersion curves are presented in Fig.~\ref{fig.ph_band}.
All investigated systems are stable in P2$_{1}$/c structure, as indicated by the absence of the soft modes.
Introduction of the non-analytical term leads to the negligible small LO-TO splitting (not shown).
Increase of the alkali metal mass leads to shift of the large part of the phonon bands to lower frequencies (cf. panels from the top to the bottom for series KBi$\rightarrow$RbBi$\rightarrow$CsBi).
Independently of this, in the phonon band structure of KBi a specific group of a few branches can be distinguished (in the range from $\sim 1.5$~THz to $\sim 2.1$~THz; marked by the green band), which is separated from the rest of the spectrum.
Analogously, similar groups (marked by yellow bands) can be found at high frequencies for RbBi and CsBi (above $2.5$~THz for both compounds).
Analysis of the partial phonon density of states (Fig.~\ref{fig.ph_dos}) uncovers the character of these branches.
In the case of KBi (RbBi and CsBi), the selected group of the bands corresponds to vibrations associated  with K (Bi) atoms.
Moreover, these groups of bands for RbBi and CsBi mainly correspond to the vibrations of Bi chains along the ${\bm b}$ ($\parallel y$) direction (cf. Fig.~\ref{fig.schemat}).

%%%%%%%%%%%%%%%%%%%%%%%%%%%%%%%%%%%%%%%
%%%%%%%%%%%%%%%%%%%%%%%%%%%%%%%%%%%%%%%
%%%%%%%%%%%%%%%%%%%%%%%%%%%%%%%%%%%%%%%
%%%%%%%%%%%%%%%%%%%%%%%%%%%%%%%%%%%%%%%

\subsection{Selection rules and Raman modes}
\label{sec.raman}

At the $\Gamma$ point, the phonon modes of $A$Bi compounds can be decomposed into the irreducible representation of the space group P2$_{1}$/c as follows:
\begin{eqnarray}
\nonumber \Gamma_\text{aoustic} &=& A_\text{u} + 2 B_\text{u} , \\
\nonumber \Gamma_\text{optic} &=& 24 A_\text{g} + 23 A_\text{u} + 24 B_\text{g} + 22 B_\text{u} .
\end{eqnarray}
In total, there are 96 non-degenerated vibrational modes.
Here, $A_\text{u}$ and $B_\text{u}$ modes are infra-red-active, while  $A_\text{g}$ and $B_\text{g}$ modes are Raman-active.

The non-resonant Raman scattering intensity depends on a direction of the incident and scattered light relatively to the principal axes of the crystal.
The intensity of the scattering is given as~\cite{loudon.01}:
\begin{eqnarray}
\label{eq.raman} I \propto | e_{i} \cdot R \cdot e_{s} |^{2} ,
\end{eqnarray}
where $e_{i}$ and $e_{s}$ are the polarization vectors of the incident and scattered light.
In the case of the P2$_{1}$/c group, the Raman tensor is given as:
\begin{eqnarray}
\label{eq.tensors} 
R \left( A_\text{g} \right) = & \left( \begin{array}{ccc}
a & d & 0 \\ 
d & b & 0 \\ 
0 & 0 & c
\end{array} \right); \quad & R \left( B_\text{g} \right) = \left( \begin{array}{ccc}
0 & 0 & 0 \\ 
0 & 0 & f \\ 
0 & f & 0
\end{array} \right) .
\end{eqnarray}
In the back-scattering configuration, $e_{i}$ and $e_{s}$ can be chosen in arbitrary planes (i.e., $xy$, $xz$, or $yz$ planes).
For study of the phonon modes propagating along the Bi atom chains in ${\bm b} \parallel y$ direction, we chose the perpendicular $xz$ plane.
In this case, the polarization vectors for linearly polarized light in the $x$ and $z$ directions are $e_{x} = (1\;0\;0)$ and $e_{z} = (0\;0\;1)$, respectively.
Similarly, the polarization vectors for left $\sigma^{+}_{xz}$ and right $\sigma^{-}_{xz}$ circularly polarized light are $\sigma_{xz}^{\pm} = \sfrac{1}{\sqrt{2}} (1\;0\;\pm i)$. 
Using relation (\ref{eq.raman}) and Raman tensors (\ref{eq.tensors}), we can determine the selection rules and Raman intensities for various scattering geometries and configurations collected in Tab.~\ref{tab.ramanselect}.

\begin{table}[!t]
\caption{
\label{tab.ramanselect}
Selection rules for the Raman-active modes of $A$Bi with P2$_{1}$/c symmetry for different configurations and backscattering geometries.
}
%\begin{ruledtabular}
\begin{tabular}{lcc}
scattering geometry & $A_\text{g}$ & $B_\text{g}$ \\
\hline 
\multicolumn{3}{c}{configuration $xy$} \\
\hline
$e_{x}$ in $e_{x}$ out (linear $\parallel$) & $|a|^{2}$ & $0$ \\
$e_{x}$ in $e_{y}$ out (linear $\perp$) & $|d|^{2}$ & $0$ \\
$\sigma^{+}_{xy}$ in $\sigma^{+}_{xy}$ out (co-circular) & $2|d|^{2}$ & $0$ \\
$\sigma^{+}_{xy}$ in $\sigma^{-}_{xy}$ out (cross-circular) & $2|a|^{2}$ & $0$ \\
\hline
\multicolumn{3}{c}{configuration $xz$} \\
\hline 
$e_{x}$ in $e_{x}$ out (linear $\parallel$) & $|a|^{2}$ & $0$ \\
$e_{x}$ in $e_{z}$ out (linear $\perp$) & $0$ & $0$ \\
$\sigma^{+}_{xz}$ in $\sigma^{+}_{xz}$ out (co-circular) & $\sfrac{1}{2}|a-c|^{2}$ & $0$ \\
$\sigma^{+}_{xz}$ in $\sigma^{-}_{xz}$ out (cross-circular) & $\sfrac{1}{2}|a+c|^{2}$ & $0$ \\
\hline
\multicolumn{3}{c}{configuration $yz$} \\
\hline 
$e_{y}$ in $e_{y}$ out (linear $\parallel$) & $|a|^{2}$ & $0$ \\
$e_{y}$ in $e_{z}$ out (linear $\perp$) & $0$ & $|f|^{2}$ \\
$\sigma^{+}_{yz}$ in $\sigma^{+}_{yz}$ out (co-circular) & $\sfrac{1}{2}|b-c|^{2}$ & $2|f|^{2}$ \\
$\sigma^{+}_{yz}$ in $\sigma^{-}_{yz}$ out (cross-circular) & $\sfrac{1}{2}|b+c|^{2}$ & $0$
\end{tabular}
%\end{ruledtabular}
\end{table}

As we can see, it is possible to distinguish Raman active modes $A_\text{g}$ and $B_\text{g}$.
Additionally, in the cases of the $xz$ and $yz$ configurations, intensity obtained for $A_\text{g}$ modes within the co-circular and cross-circular scattering geometries are different.
In contrast to this, for $B_\text{g}$ modes, only co-circular scattering geometry gives non-zero insensitivity.
A distinction between $A_\text{g}$ modes at the $\Gamma$ point in co-circular and cross-circular scattering geometries gives information about the possible presence of the chiral modes in the system far from the $\Gamma$ point.
Indeed, this is allowed in the presence of rotational symmetry.
Moreover, these properties of the $A_\text{g}$ mode can be used as a tool for experimental confirmation of chiral modes existence~\cite{zhao.xu.22}.
For example, in the case of the 2D transition metal dichalcogenides~\cite{chen.zheng.15,zhu.yi.18,chen.lu.19,zhang.srivastava.20}, the circular polarized Raman spectra can ``interplay'' with the valleys of the electronic band structure~\cite{drapcho.kim.17,tatsumi.saito.18}, due to the selection rules.
Similarly, in the case of Fe$_{3}$GeTe$_{2}$~\cite{du.tang.19}, an unbalance in the Raman scattering peaks for circular polarized light is observed, which is possible due to the excitation of chiral modes by the light with different circulation.
Similar configuration was used in the study of CrBr$_{3}$~\cite{yin.ulman.21}.
And finally, in the case of chiral system $\alpha$-HgS~\cite{ishito.mao.21}, the shift of the anti-Stokes and Stokes Raman spectra for co-circular and cross-circular configuration can be observed.
This is allowed when the chiral modes are present in the vicinity of the $\Gamma$ point.

%%%%%%%%%%%%%%%%%%%%%%%%%%%%%%%%%%%%%%%
%%%%%%%%%%%%%%%%%%%%%%%%%%%%%%%%%%%%%%%
%%%%%%%%%%%%%%%%%%%%%%%%%%%%%%%%%%%%%%%
%%%%%%%%%%%%%%%%%%%%%%%%%%%%%%%%%%%%%%%
%%%%%%%%%%%%%%%%%%%%%%%%%%%%%%%%%%%%%%%

\begin{figure}{!pt}
\centering
\includegraphics[width=0.75\linewidth]{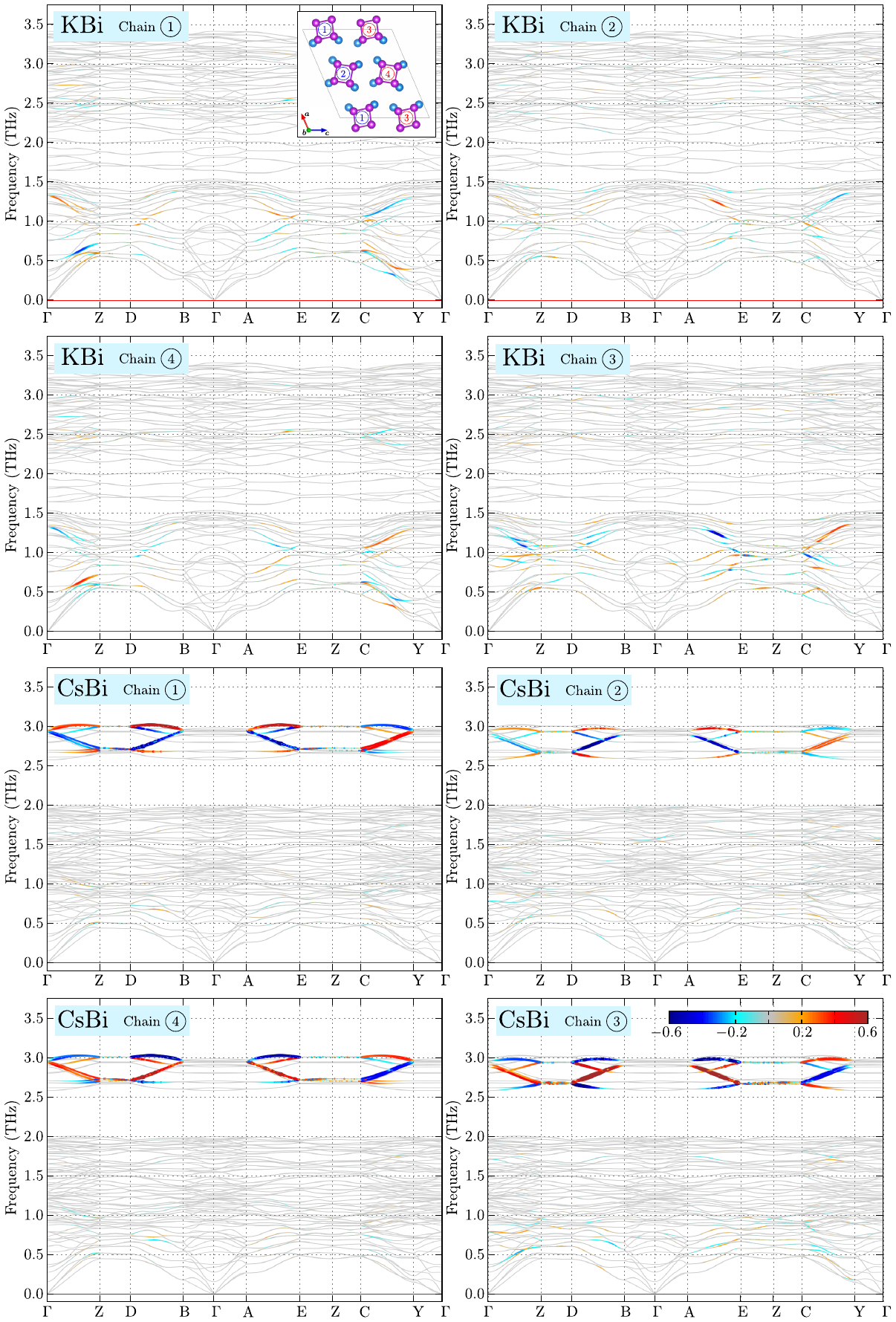}
\caption{
Total phonon polarization of the chosen chains (as labeled) for KBi and CsBi. 
The numbers of the chains are presented in insets.
\label{fig.chiral}
}
\end{figure}

\subsection{Chiral phonons in clean system}
\label{sec.chiral_clean}

At the high symmetry points $\Gamma$, K, and K', in the hexagonal system with three-fold rotational symmetry, the PAM $l_\text{PAM}$ of phonons can be defined as~\cite{zhang.niu.14,zhang.niu.15}:
\begin{eqnarray}
C_{3} \; \text{e} = \exp \left( - i \frac{2\pi}{3} l_\text{PAM} \right) \text{e} , 
\end{eqnarray}
where $\text{e}$ is the wave function (polarization vector; see Sec.~\ref{sec.chiral_teo}) of the phonon with wavenumber ${\bm q}$.
In this formulation, $l_\text{PAM}$ can have values of $-1$, $0$, or $1$.
In the presence of the inversion symmetry and the time reversal symmetry, the PAM of the system can be zero, even in the presence of the chiral phonons~\cite{coh.19}.
Typically, in such a case, each atom exhibiting the left-handed circular motion around the equilibrium position has a partner with right-handed circular motion, and {\it vice versa}. 
However, even the simple breaking of the spatial inversion symmetry, e.g., by introducing a staggered sublattice on-site potential or isotope doping, can lead the non-zero PAM of the system~\cite{zhang.niu.15}.

The above PAM definition can be extended to the non-symmorphic groups~\cite{zhang.murakami.21}.
In fact, existence of the rotational symmetry in the system guarantees presence of the chiral phonons.
In such a case, the PAM can be decomposed into a nonlocal contribution (Bloch-phase part, called also intercell/orbital part) from the Bloch phase factor, and a local contribution (self-rotation part, called also intracell/spin part) from relative vibrations between the sublattices.
Moreover, for the nonsymmorphic system, the PAM has unique features, e.g., can have noninteger value from the fractional-translational part of the screw rotation symmetry.
Nevertheless, the PAM can be expressed also in the phonon circular polarization, introduced in Sec.~\ref{sec.chiral_teo}, and it is closely related to the circular motion of atoms.

In  Fig.~\ref{fig.chiral} we present the phonon circular polarization calculated for chosen chains of KBi and CsBi materials (the results for RbBi, not shown, are qualitatively the same as for CsBi, cf. Fig.~\ref{fig.ph_band}).
As we can see, in the case of KBi the chiral phonons can be found for  low-frequency branches. 
However, these mods are strongly ``hybridized'' with the ordinary modes.
In contrast, for the case of CsBi (and RbBi) the chiral modes are present in the range of frequencies, which correspond to vibrations only within the Bi atoms chains (i.e. high frequencies).
In investigated cases, the phonon circular polarization for each atom and chain has an absolute value smaller than nominal (i.e., equal $1$), which suggests that the atoms move along elliptical orbits around the equilibrium position instead of perfect circles.
Indeed, the Supplemental Material (SM)~
\footnote{ %==============##########
See Supplemental Material available online for visualizing the chiral modes in described systems.
Movie \texttt{DBp2\_modeXX.mp4} presents vibrations generated by the \texttt{XX} mode for the (D+B)/2 wave vector.
Movie \texttt{Zp10\_mode\_propagation.mp4} presents propagation of the chiral modes along the Bi atoms chains for wave vector Z/10 generated by the 93$^\text{th}$ mode.
} %===============###########
contains movies \texttt{DBp2\_modeXX} illustrating the vibrations caused by the chiral modes (where \texttt{XX} denotes No. of the mode) with wave vector (D+B)$/2$.
Additionally, the total phonon circular polarization is equal zero.
This is associated with existence of  four chains with opposite chirality.
For example, when the propagating chiral mode causes the left-handed circulation in the chain \circled{1}, the right-handed circulation occurs in the chain \circled{4} at the same time.

\subsection{Chiral modes with non-zero pseudoangular momentum induced by impurities}
\label{sec.chiral_imp}

\begin{figure}[!t]
\centering
\includegraphics[width=0.75\linewidth]{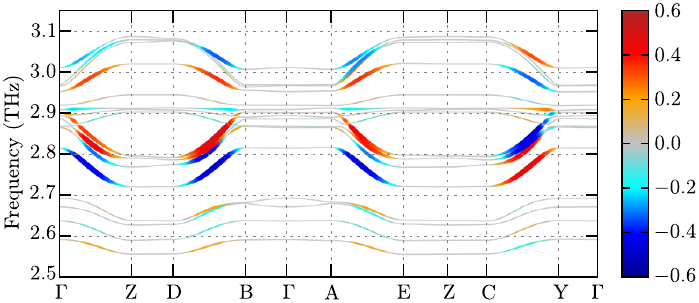}
\caption{
Chiral phonons induced in CsBi induced by the Rb impurity (i.e., Cs$_{1-x}$Rb$_{x}$Bi for $x=1/16$).
Color corresponds to the total phonon circular polarization of the system.
\label{fig.doping}
}
\end{figure}

From a practical point of view, it is important to have a system with the non-zero PAM.
Unfortunately, as we mention earlier, our (clean) system is characterized by the vanishing total phonon circular polarization (i.e., PAM equivalent).
Here it is worth to mention, that the PAM can be induced in the system by the symmetry breaking, which is reflected in the IFC modification~\citep{coh.19}.
In this context, it is worth to notice  that investigated compounds are isostructural during K$\rightarrow$Rb$\rightarrow$ Cs substitution. 
It can be assumed that the system has small concentration of the alkali metal impurities.
Such doping, naturally, leads to symmetry breaking.
In our investigation, we substituted Cs atom by Rb atom within the CsBi crystal.
Here, we present direct calculations for the case of Cs$_{1-x}$Rb$_{x}$Bi with $x=1/16$, i.e., with one Rb impurity atom in $2 \times 1 \times 2$ supercell of CsBi.
During the calculations, all symmetries of the system are lifted, and the system symmetry is reduced to P1 group (this was found from the symmetry analysis after optimization of the supercell within the standard DFT procedure).
Nevertheless, the supercell still contains the main structure of the homogeneous system, presented in Fig.~\ref{fig.schemat}.
For such optimized structure, we calculate the total phonon circular polarization (the branches exhibiting the chiral phonons are shown in Fig.~\ref{fig.doping}).

Introduction of the impurity to the system leads to removal of part or all of the symmetry (in our case all symmetries), and the band decoupling (cf. bottom panels of Fig.~\ref{fig.ph_band} and Fig.~\ref{fig.doping}).
Similar decoupling of the bands can be realized by the external magnetic field~\cite{wang.li.21} or strain~\cite{ptok.kobialka.21}.
As we can see (Fig.~\ref{fig.doping}), due to the absence of the symmetries, all bands have lifted degeneracy.
However, in the system prepared in such a way the non-zero total phonon circular polarization (and equivalently the non-zero total PAM) is observed.
This property of the system can be used in practice to the intentional (by-hand) induced chiral phonons.
Additionally, the chiral phonons are realized in the vicinity of the $\Gamma$ point, which can be useful in experimental confirmation of our prediction.

Described above case with induced  chiral phonon with the non-zero total PAM is similar to well-know realization of the chiral phonons in the honeycomb lattice~\cite{zhang.niu.15}.
In the ideal case, i.e., of the homogeneous system, due to the inversion symmetry and the time reversal symmetry, the total PAM is zero. 
However, introduction of different atomic masses in two non-equivalent sublattice of the honeycomb lattice leads to emergence of chiral phonons with the non-zero total PAM.

As we mentioned earlier, the $A$Bi$_{2}$ compounds with P2$_{1}$/c posses the mirror symmetry $\{ m_{010} | 0 \; \sfrac{1}{2} \; \sfrac{1}{2} \}$, and the two-fold rotational symmetry $\{ 2_{010} | 0 \; \sfrac{1}{2} \; \sfrac{1}{2} \}$ (both symmetries are nonsymmorphic).
Realization of such symmetries  is reflected on the symmetry within the reciprocal space [the mirror plane and the rotational axis are presented schematically in Fig.~\ref{fig.mode}(a)] and in the features of the realized chiral modes [Fig.~\ref{fig.doping}].
Observed features are completely different from these reported in the systems with hexagonal symmetry, where the chiral modes with non-zero PAM were observed in the high symmetry points with different valleys (K and K' points).
Simultaneously, as a consequence of the full winding of the wavefunction phase, along the path between two $\Gamma$ points from the neighboring Brillouin zones,  the chiral modes disappear at some high-symmetry point (in the case of the system with the hexagonal symmetry, they are observed in the M point).
In our case, this is related to the all points at the Brillouin zone edges.
In practice, the PAM of chiral modes vanishes at the high symmetry points, what is guarantee by the two-fold rotational symmetry.
However, the chiral modes can be still expected along some paths between the high symmetry points [Fig.~\ref{fig.doping}].

\subsection{Propagation of chiral modes}
\label{sec.chiral_prop}

\begin{figure}
\centering
\includegraphics[width=0.75\linewidth]{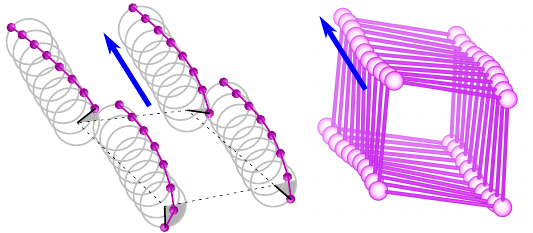}
\caption{
Representation of the two-fold rotational symmetry axis and the mirror plane in the reciprocal space (a).
Cartoon presenting chiral mode propagation along the Bi chain (b).
The black dashed quadrangle represents equilibrium positions of atoms.
Example of the Bi atoms displacement within one of the four Bi chains caused by the $93^\text{th}$ mode at \mbox{(D+B)$/2$} point in CeBi (c).
The blue arrow corresponds to the $y$ axis parallel to the chains.
\label{fig.mode}
}
\end{figure}

Now, we discuss the propagation of the chiral modes along the Bi atoms chains.
The primitive cell contains four chains, two with left-handed and two with right-handed chiralities.
Each chain (in the primitive cell) is constructed by four Bi atoms, creating square-like structure in the ${\bm a}{\bm c}$ plane (see Fig.~\ref{fig.schemat}), and they can be treated as a separate sublattices.
For our analysis, we chose the $93^\text{th}$ (chiral) mode with wave vector Z/10.
Propagation of this mode is presented in the \texttt{Zp10\_mode\_propagation} movie in SM, and also schematically in Fig.~\ref{fig.mode} (the left panel presents a cartoon of the mode propagation, while the right panel shows displacement found in the  calculations).
Due to the different positions of the atoms in space, the difference in the initial phase is visible (represented by the grey triangles in Fig.~\ref{fig.mode}).
During the mode propagation, this phase is also propagating along the chains and the characteristic screw movement along $y$ axis (blue arrow) is observed.
However, we should have in mind, that the presence of chiral modes is related to the motion of the atoms over elliptic orbits (and the circular polarization of the phonon is a measure of this). 
The ``full'' $2\pi$ phase in redistributed over all atoms forming the chain (related to the given wave number). 
For a chosen mode, the period of the propagating wave is given as $2\pi/4 \mathcal{N}$, where $\mathcal{N}$ denotes the number of  cells along the chain direction.
Existence of the chiral modes propagating along one direction opens a new way for application of chiral systems in the chirality-based quantum devices~\cite{chen.wu.21a}, due to  the transport of the quantized chirality information as well as angular momentum through the crystal. 
Such propagation process is ``protected'' by the (pseudo)angular momentum conservation law~\cite{tatsumi.kaneko.18,ruckriegel.stereib.20}.
A situation can be similar to the propagation of the chiral phonons in the oxide heterostructures~\cite{jeong.kim.22}, where the chiral phonons can dynamically mediate the long-range magnetic interaction along well-coupled layers of the system.
In our case, we can speak of coupling between Bi atoms forming separated chains.
Additionally, two channels of propagation are allowed, due to the existence of different types of the chains (with different chirality).

\AP{
\paragraph*{Possible applications ---}
Similarly like in the spintronics~\cite{zutic.fabian.04}, where the active control and manipulation of spin degrees of freedom in solid-state systems are used, one can introduce the {\it phonontronics} term~\cite{li.ren.12}, where the features of the phononic system can be used.
%%%
From the application point of view, systems which possess non-zero PAM are the most important.
As a consequence of the selective rules~\cite{tatsumi.kaneko.18,ruckriegel.stereib.20}, such materials could develop ways for transferring the PAM from photons to electron spins via propagating chiral phonons in future devices.
%%%
For example, the propagating chiral phonons can transport quantized chirality information as well as PAM through the crystal lattice~\cite{chen.wu.21a}.
In such context, the most promising are the chiral systems like $\alpha$-HgS~\cite{ishito.mao.21}, which also allow for  realization of the diode-like effect~\cite{chen.wu.22}.
Propagating such chiral phonons along the system can affect on the interlayer exchange coupling due to the strong the chiral phonon--spin coupling, and such phenomenon was observed experimentally~\cite{jeong.kim.22}.
%%%
In the context of the possible application, one should also mention the phonon Hall effect~\cite{strohm.rikken.05}.
Such phenomena are interesting in a context of eventual applications of observed thermal Hall effect (which, in contrary to the phonon Hall effect, is mostly associated with the magnons propagation) in $\alpha$-RuCl$_{3}$~\cite{kasahara.ohnishi.18,czajka.gao.23}.
Moreover, the interplay between the chiral phonons transport and topological magnons can play an important role~\cite{thingstad.kamra.19}.
%%%
Finally, the chiral phonon activated spin Seebeck effect has been observed recently~\cite{kim.vetter.23} in the materials without magnetic order nor spin--orbit coupling, what gives new opportunities for designing advanced spintronic devices based on nonmagnetic chiral materials.
}

%%%%%%%%%%%%%%%%%%%%%%%%%%%%%%%%%%%%%%%
%%%%%%%%%%%%%%%%%%%%%%%%%%%%%%%%%%%%%%%
%%%%%%%%%%%%%%%%%%%%%%%%%%%%%%%%%%%%%%%
%%%%%%%%%%%%%%%%%%%%%%%%%%%%%%%%%%%%%%%
%%%%%%%%%%%%%%%%%%%%%%%%%%%%%%%%%%%%%%%

\section{Summary}
\label{sec.summary}

This paper discusses basic properties of the $A$Bi ($A=$K, Rb, Cs) binary compounds.
These compounds crystallize in P2$_{1}$/c structure, containing both left-handed and right-handed chiral chains of Bi atoms. 
The electronic band structure exhibits insulating character with a gap of $\sim 0.25$~eV.
The phonon dispersion curves have no imaginary frequencies, which indicates that all compounds are stable in the P2$_{1}$/c  structure.
The group theory analysis of phonon modes at $\Gamma$ point shows only non-degenerate modes, which can be distinguished with the Raman spectroscopy measurements.

Our calculation shows that chiral phonons can be realized in the vicinity of the $\Gamma$ point within the system which does not feature three-fold rotational symmetry. 
In case of RbBi and CsBi the chiral phonon modes are separated from the other modes in the phonon spectra.
The chiral modes are restricted only to the Bi atoms chains.
We have found screw-like movement of atoms around their equilibrium positions as the chiral modes propagate along the chains.
At the same time, the phase of the atom displacement is 'transferred' along the chains.
We also discuss chiral modes induced by the impurities in the system.
In our case, for CsBi doped by Rb, we show that the chiral modes with a non-zero pseudoangular momentum can be induced as a result of the transnational symmetry breaking.
Then, similarly to the case of the time reversal symmetry breaking, the decoupling of all chiral modes is observed.
This opens new ways for realization of the phonon chiral modes.
Due to the selection rules, the existence of these chiral modes can be relatively simply detected in the Raman scattering experiments with co-circular or cross-circular polarized light.

\section*{Acknowledgments}

Some figures and movies in this work were rendered using 
{\sc Vesta}~\cite{momma.izumi.11} and {\sc VMD}~\cite{vmd} software.
J.S.~acknowledges the hospitality of the Henryk Niewodnicza\'{n}ski Institute of Nuclear Physics of the Polish Academy of Sciences in Krak\'{o}w during his student internship.
This work was supported by the National Science Centre (NCN, Poland) under grants 
2017/24/C/ST3/00276 (K.J.K.),
2017/25/B/ST3/02586 (P.T.J.)
and
2016/21/D/ST3/03385 (A.P.).
K.J.K.~thanks the Polish National Agency for Academic Exchange for funding in the frame of the Bekker programme (PPN/BEK/2020/1/00184).
In addition,  K.J.K. and A.P. are grateful for the funding from the scholarships of the Minister of Science and Higher Education (Poland) for outstanding young scientists (2019 edition, Nos.~821/STYP/14/2019 and 818/STYP/14/2019, respectively).

\section*{Declaration of Competing Interest}

The authors declare that they have no known competing financial interests or personal relationships that could have appeared to influence the work reported in this paper.

\section*{Data Availability Statement}

The data presented in this study are available on request from the authors.

%%%%%%%%%%%%%%%%%%%%%
%%%%%%%%%%%%%%%%%%%%%
%%%%%%%%%%%%%%%%%%%%%

%\nocite{*}

\bibliographystyle{spphys}
\bibliography{biblio.bib}

\end{document}